\documentclass{aa}

\usepackage{natbib}
\usepackage{graphicx}
\usepackage{amssymb}
\usepackage[dvipsnames]{color} 

\begin{document}

\def\h{H~I}
\def\he{He~I}
\def\hei{He \,II}
\def\12{{1\over 2}}
\def\msun{M_{\odot}}
\def\lsun{L_{\odot}}
\def\div{\nabla\cdot}
\def\grad{\nabla}
\def\rot{\nabla\times}
\def\eg{{\it e.g.,~}}
\def\ie{{\it i.e.,~}}
\def\etal{{\it et~al.,~}}
\def\de{\partial}
\def\ltsima{$\; \buildrel < \over \sim \;$}
\def\simlt{\lower.5ex\hbox{\ltsima}}
\def\gtsima{$\; \buildrel > \over \sim \;$}
\def\simgt{\lower.5ex\hbox{\gtsima}}
\def\noi{\noindent}
\def\bs{\bigskip}
\def\ms{\medskip}
\def\ss{\smallskip}
\def\ob{\obeylines}
\def\l{\line}
\def\hrf{\hrulefill}
\def\hf{\hfil}
\def\q{\quad}
\def\qq{\qquad}
\def\Red#1{\color{red}{#1}}

\newcommand{\be}{\begin{equation}}
\newcommand{\ee}{\end{equation}}
\newcommand{\msunpctwo}{\mbox{\, M$_{\sun} \, {\rm pc}^{-2}$}}
\newcommand{\natd}[2]{\mbox{$#1 \cdot 10^{#2}$}}
\newcommand{\pder}[2]{\frac{\partial #1}{\partial #2}}
\newcommand{\pdert}[1]{\pder{#1}{t}}
\newcommand{\mcorr}{\bf}         
\newcommand{\mcorrm}{\boldmath}  
\newcommand{\bl}[1]{\mbox{\boldmath$ #1 $}}


\title{The age of blue LSB galaxies}
\titlerunning{Age of blue LSB galaxies} 
\authorrunning{Vorobyov et al.}
 
\author{E. I. Vorobyov\inst{1,2},  Yu. Shchekinov 
\inst{5}\thanks{on leave from Department of Physics, South Federal  University, and 
Special Astrophysical Observatory}, and D. Bizyaev \inst{3,4}, 
D. Bomans\inst{5}, R.-J. Dettmar\inst{5},  }

\offprints{E. I. Vorobyov}
\institute{The Institute for Computational Astrophysics, Saint Mary's University, Halifax NS, 
B3H 3C3, Canada 
\and
Institute of Physics, South Federal
University, Stachki 194, Rostov-on-Don, Russia 
\and
New Mexico State University/APO, Sunspot, NM, 88349, USA  
\and
Sternberg Astronomical Institute, Universitetsky 13, Moscow, 119992, Russia
\and 
Astronomisches Institut, Ruhr-Universit\"at Bochum, Bochum, Germany
}

\date{} 

\abstract
{Low metallicities, large gas-to-star mass ratios, and blue colors of most 
low surface brightness (LSB) galaxies imply that these systems may be younger than their
high surface brightness counterparts.}
{We seek to find observational signatures that can help to constrain the age 
of blue LSB galaxies.}
{We use numerical hydrodynamic modelling to study the long-term ($\sim 13$~Gyr) 
dynamical and chemical
evolution of blue LSB galaxies adopting a sporadic scenario for star formation. 
Our model galaxy consists of a thin gas disk, stellar disk with a developed spiral structure 
and spherical dark matter halo. Our models utilize various rates of
star formation and different shapes of the initial mass function (IMF). 
We complement hydrodynamic modelling with population synthesis modelling to produce 
the integrated $B-V$ colors and H$\alpha$ equivalent widths.}
{We find that the mean oxygen abundances, $B-V$ colors, H$\alpha$ equivalent widths, 
and the radial fluctuations
in the oxygen abundance, when considered altogether, can be used to constrain the age of blue 
LSB galaxies if some independent knowledge of the IMF is available. 
Our modelling strongly suggests the existence of a minimum age for 
blue LSB galaxies. Model $B-V$ colors and mean oxygen abundances 
set a tentative minimum age at 1.5--3.0~Gyr, whereas model H$\alpha$ equivalent widths
suggest a larger value of order 5--6~Gyr. The latter value may decrease somewhat, if blue LSB
galaxies host IMFs with a truncated upper mass limit. We found no firm evidence that 
the age of blue LSB galaxies is significantly smaller than 13 Gyr. }
{The age of blue LSB galaxies may vary between 1.5--6~Gyr and 13~Gyr, depending on 
the physical conditions in the disk that control the form of the IMF and the rate of star formation.
A failure to observationally detect 
large radial fluctuations in the oxygen abundance 
of order 0.5--1.0~dex, which, according to our modelling, are characteristic of (1-2)-Gyr-old 
galaxies, will argue in favour of the more evolved nature of blue LSB galaxies.}
\keywords{ISM:galaxies -- galaxies:abundances -- galaxies:evolution}

\maketitle

 

\section{Introduction }
Low surface brightness (LSB) galaxies are the galaxies whose central surface brightness is much 
fainter than the Freeman value $\mu_{\rm B}=21.65 \pm 0.30$~mag~arcsec$^{-2}$.
These galaxies are thought to represent a significant fraction of the galaxy number density 
in the universe \citep{mcg95,oneil00,trach06}. LSB galaxies show a wide spread of 
colours ranging from red ($B-V\simeq 1.7$) to very blue ($B-V\simeq 0.2$) \citep{oneil},
but the most common type seems to be blue LSB galaxies: late-type, disk dominated spirals with central
surface brightness $\mu_{\rm B}\ga 23$~mag~arcsec$^{-2}$ and colors lying in the range $(B-V)=0.3-0.7$.
These blue LSB galaxies are among the most gas-rich (up to 50\% of the total baryonic mass) 
and metal-deficient (about 5\%-20\% of the solar metallicity) galaxies 
\citep{mcg94,roen95,mcg97,deBlok98,Kuzio04}.  

Low metallicities and large gas-to-star mass ratios may be interpreted in various ways. 
For instance, blue LSB galaxies may have formed recently, at redshifts close to 
$z=0$. However, this can confront the hierarchical scenario of large-scale structure 
formation, within which low-mass galaxies form first at redshifts close 
to the onset of the reionization epoch, $z\sim 10$. Therefore, it is more likely that
blue LSB galaxies have formed at large $z$ but have had a delayed onset of major
star formation epoch.  In this case, a conflict with the hierarchical structure formation is relaxed.
Indeed, blue LSB galaxies show a wide range of masses with 
an excess of dwarf systems with luminosities of order $10^7-10^8\lsun$  \citep{spray}. 
Such dwarf systems are known to have a younger mean stellar age than their more massive 
counterparts -- a phenomenon known as `downsizing' \citep[e.g.][]{Thomas}.
Another possible interpretation of the apparent unevolved nature of blue LSBs is that
they do not form late, nor have a delayed onset of star formation, but simply evolve
slower than their high surface brightness counterparts \citep[e.g.][]{Hoek}.

There have been many efforts to infer the age of blue LSB galaxies from 
the properties of stellar population. 
Broadband photometric studies, complemented by H$\alpha$ emission line data 
and synthetic stellar population code modeling, predict quite a wide range for the ages of 
blue LSB galaxies: from 1-2~Gyr \citep{zack06} to 7-9~Gyr \citep{padoan,jimenez}. 
Rather young ages of blue LSB galaxies ($<5$ Gyr) follow also from their $(V-I)$ colors 
and HI content \citep{schom}. The comparison of measured spectral energy distributions in a sample 
of LSB galaxies with the synthetic spectra (derived assuming an exponentially declining
SFR) by \citet{Haberzettl} suggests the ages of the dominant stellar population
between 2 and 5 Gyr.

In this paper, we perform numerical hydrodynamic modelling of the dynamical, 
chemical, and photometric evolution of blue LSB galaxies with the purpose to constrain 
their typical range of ages.
The cornerstone supposition of our modelling, motivated by observational evidence, 
is a sporadic nature of star formation in blue LSB galaxies.
Indeed, \citet{mcg94}, \citet{deBlok95}, \citet{Gerretsen}, and others argue that 
the primary cause for the blue colors is the young age of the dominant stellar 
population. This can take the form of young local bursts of star formation superimposed (or not) 
on a continuous low-rate star formation. Available H$\alpha$ images of LSB galaxies  
\citep[see e.g.][]{McGaugh1994,McGaugh1995,Auld} appear to support this so-called sporadic star formation
scenario.  Current star formation is localized to a handful of
compact regions. There is little or no diffuse H$\alpha$ emission coming
from the rest of the galactic disk. Randomly distributed star formation sites, along with 
inefficient spatial stirring of heavy elements, give rise to radial fluctuations
in the oxygen abundance on spatial scales of order $1-2$~kpc. The resulted 
fluctuation spectrum is age-sensitive and can be potentially used to set {\it order-of-magnitude} 
constraints on the age of blue LSB galaxies. 
However, the age prediction may be considerably improved by using the mean oxygen abundances and H$\alpha$
equivalent widths provided that some independent knowledge of the initial mass function is available.

The paper is organized as follows. In the next Section 
we discuss qualitatively the processes that mix heavy elements in disks
of LSB galaxies. Section~\ref{modelgalaxy} 
details our theoretical model, while Section~\ref{observe} describes our 
observations. In Sections~\ref{results} and \ref{synthesis} we discuss our numerical results, and 
Section~\ref{conclude} presents main conclusions.


\section{Mixing of heavy elements in LSB disks}
\label{mix}
In high surface brightness
(HSB) galaxies mixing of heavy elements is provided by multiple shock waves from 
supernova explosions. It requires approximately 100~Myr to homogenize 
heavy elements in the warm neutral interstellar medium of HSB galaxies  \citep{deav02}. LSB 
galaxies, on the other hand, are characterized by the star formation rate ($\sim 0.1 \msun$ yr$^{-1}$)
that is almost two orders of magnitude lower than the Milky Way value, 
$\approx 5 \msun$ yr$^{-1}$. Consequently, the frequency of shock waves in the 
interstellar medium of LSB galaxies and the 
corresponding volume filling factor occupied by the shock processed gas 
can be as small as $\Psi_{sh}\simlt (200~{\rm Myr})^{-1}$ and 
$f_{sh}\simlt 0.004$, respectively. For comparison, the corresponding values
in the Milky Way are $\Psi_{sh}\sim 
(5~{\rm Myr})^{-1}$ \citep{draine79} and $f_{sh}\sim 0.2$ \citep{deav05}. 
In these conditions, homogenization of heavy elements in 
LSB galaxies may require a considerably longer time scale than 100~Myr, a typical value for 
HSB galaxies. Differential rotation 
of the galactic disk and radial convection driven by the spiral stellar
density waves may in principle be efficient in mixing of heavy elements
in LSB galaxies. However, the time scales for these processes in LSB galaxies 
are not well known and numerical simulations are needed to determine their importance. 
Therefore, the distribution 
of heavy elements in the disk of an LSB galaxy  may be characterized by
radial variations, the typical amplitude of which is expected to diminish with time. 
It is our purpose to determine whether these radial abundance fluctuations 
can be used to constrain the ages of LSB galaxies.  

We note that mixing in the azimuthal direction is strongly enhanced by 
differential rotation. Namely, the size of a mixing eddy grows with time 
due to differential rotation as $\Delta x\sim \left({u_\phi}\right)^\prime_{r} D^{1/2}t^{3/2}$, 
much faster than due to hydrodynamic 
mixing $\Delta x\propto t$ \citep{scalo}, where $\left({u_\phi}\right)^\prime_{r}$ 
is the radial derivative of 
the rotation velocity, $D$ is the diffusion coefficient. Differential rotation therefore 
homogenizes metals in the azimuthal direction faster than in the radial one and the radial
variations remain for a longer time. 

\section{Model LSB galaxy}
\label{modelgalaxy}
\subsection{Initial configuration}
\label{model}
LSB galaxies show a wide range of morphological types, ranging from giant spirals to 
dwarf irregular galaxies, from early-type red galaxies to late-type blue ones. 
In this paper we focus on late-type blue LSB galaxies.
According to the observational data \citep{Hoek}, a typical blue LSB galaxy has a 
dynamical mass of order $10^{10}~M_\odot$ and HI mass of order $2\times 10^9~M_\odot$.
We closely follow these estimates when constructing our model galaxy.
Our model LSB galaxy consists of a thin gas disk, which evolves in the external 
gravitational potential of both the spherical dark matter halo and extended stellar disk. 
LSB galaxies often feature a few ill-defined spiral arms. Therefore, we assume that our model 
stellar disk has developed a two-armed spiral structure. Below, we provide a more
detailed explanation for each of the constituents of our model galaxy.

The velocity field of most LSB galaxies is better fitted by dark matter models described by 
a modified isothermal sphere rather than by a NFW profile \citep{Kuzio}. Therefore,
the spherical dark matter halo in our model is modelled by a radial density profile described by 
the modified isothermal sphere
\begin{equation}
\rho_{\rm h}={\rho_{\rm h0}\over (1+r/r_{\rm h})^2},
\label{halodens}
\end{equation}
where $\rho_{\rm h0}=6.0 \times 10^{-3}~M_\odot$~pc$^{-3}$ and $r_{\rm h}=5.7$~kpc  
are the central volume density and characteristic scale length of the dark matter halo,
respectively, and $r$ is the radial distance. 
The adopted values of $\rho_{\rm h0}$ and $r_{\rm h}$ are similar to those
obtained by \citet{Kuzio} for UGC~5750. The dark matter halo mass in the inner 15~kpc
is approximately $2.0\times 10^{10}~M_{\odot}$.
The radial gravity force (per unit mass) due to the spherical dark matter halo in the plane of the
disk can be written as 
\begin{equation}
{\partial \Phi_{\rm h} \over \partial r}=4 \pi G \rho_{\rm h0}
r_{\rm h}\left[ r/r_{\rm h} - \arctan(r/r_{\rm h}) \right]
\left({r_{\rm h}\over r}\right)^2.
\label{halo}
\end{equation}

The stellar component of our model galaxy consists of two parts -- an axisymmetric 
stellar disk and a two-armed spiral pattern. The former is described by a radially declining 
exponential profile
\begin{equation}
\Sigma_{\rm st} = \Sigma_{\rm s0} \exp(-r/r_{\rm s}),
\label{stellar}
\end{equation}
where $\Sigma_{\rm s0}=30~M_\odot$~pc$^{-2}$ is the central stellar surface density
and $r_{\rm s}=4$~kpc is the radial scale length of the stellar disk. The adopted value for $r_{\rm
s}$ is typical for blue LSB galaxies \citep{deBlok95}.  We note that our model 
stellar disk serves as
a proxy to the real stellar disk that is expected to form during the evolution of our model galaxy.
The mass of the stellar disk in the inner 15~kpc is $M_{\rm s}=2.6\times 10^{9}~M_\odot$.
For numerical purposes, the gravitational potential of the axisymmetric thin stellar 
disk  in the plane of the disk is calculated as
\begin{equation}
\Phi_{\rm st}(r,z=0)= - \pi G \Sigma_{\rm s0} r \left[I_0(y) K_1(y) - I_1(y) K_0(y) \right],
\end{equation}
where $I_n$ and $K_n$ are modified Bessel functions of the first and second kinds, respectively,
and $y\equiv r/(2 r_{\rm s})$.

The non-axisymmetric gravitational potential of stellar spiral arms  is described in
polar coordinates ($r,\phi$) by a running density wave \cite[see
e.g.][]{VS}
\begin{equation}
\Phi_{\rm sp}(r,\phi)=-C(r) \cos\left[ m(\cot(i) \ln(r/r_{\rm sp})+\phi - \Omega_{\rm sp} t)\right],
\label{nonsym}
\end{equation}
where $C(r)$ is the radially varying amplitude, $i$ is the pitch angle, $r_{\rm sp}$ is the characteristic radius of the spiral 
at $\phi=0$, 
$m=2$ is the number of spiral arms, and $\Omega_{\rm sp}$ is the
angular velocity of the spiral pattern.
In the following we adopt $r_{\rm sp}=6$~kpc and $\Omega_{\rm sp}=5.5$~km~s$^{-1}$~kpc. 
The pitch angle is set to $25^{\circ}$, typical for the late-type spiral
galaxies. 

\begin{figure}
  \resizebox{\hsize}{!}{\includegraphics{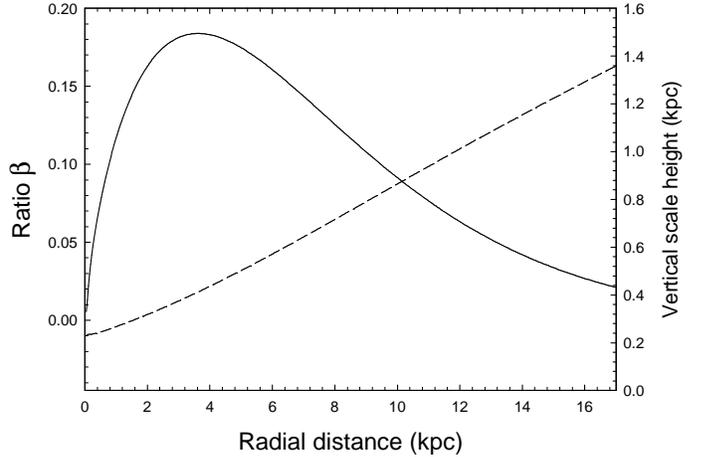}}
      \caption{ ({\it Solid line.}) Ratio $\beta(r)$ of the maximum 
non-axisymmetric gravitational acceleration due to spiral arms
$[({\partial \Phi_{\rm sp}/\partial r})^2 +(r^{-1}\,
{\partial \Phi_{\rm sp} / \partial \phi})^2]^{1/2}$  
vesus gravitational acceleration due to both the dark matter halo 
$\partial\Phi_{\rm h}/\partial r$ and axisymmetric stellar disk  
$\partial \Phi_{\rm st}/\partial r$. ({\it Dashed line.}) Vertical scale height $Z$ of our model 
gas disk as a function of radius in the beginning
of numerical simulations ($t=0$).  }
         \label{fig1}
\end{figure}

The amplitude $C(r)$ of the spiral stellar gravitational potential $\Phi_{\rm sp}$ determines the response
of the gas and consequently the appearance of a spiral pattern in the gas disk.
We adopt the following expression for the amplitude $C(r)=C_0(r)^{\alpha(r)}$. 
Here, $C_0(r)$ is a linear function of radius which has a value of 0 at $r=0$~kpc and 
equals $1.8\times10^{-3}$ 
(in dimensionless units) at $r=17$~kpc. The amplitude is set to zero at the galactic 
center to ensure that the gravitational potential of the stellar spiral density wave diminishes 
at small galactocentric distances.
Otherwise, even a small-amplitude spiral gravitational potential would produce 
strong azimuthal gravitational forces near the galactic center due to 
converging radial gridlines in polar coordinates, which quickly destroys the gas disk during 
the numerical simulations.
The exponent $\alpha(r)$ decreases linearly with radius $r$. More specifically,
we choose $\alpha(r=0~{\rm kpc})=2.3$, and $\alpha(r=17~{\rm kpc})=0.6$.
This specific form of $C(r)$ allows us to produce spiral gravitational 
potentials with different shapes and maximum amplitudes. 
The resulted ratio $\beta(r)$ of the maximum 
non-axisymmetric gravitational acceleration due to spiral arms
$[({\partial \Phi_{\rm sp}/\partial r})^2 +(r^{-1}\,
{\partial \Phi_{\rm sp} / \partial \phi})^2]^{1/2}$ (at a given radial distance $r$) 
versus the gravitational acceleration due to both the dark matter halo 
$\partial\Phi_{\rm h}/\partial r$ and axisymmetric stellar disk 
$\partial \Phi_{\rm st}/\partial r$ is shown in Fig.~\ref{fig1} by the solid line. 
We note that $\beta$ never exceeds $19\%$.

\begin{figure}
  \resizebox{\hsize}{!}{\includegraphics{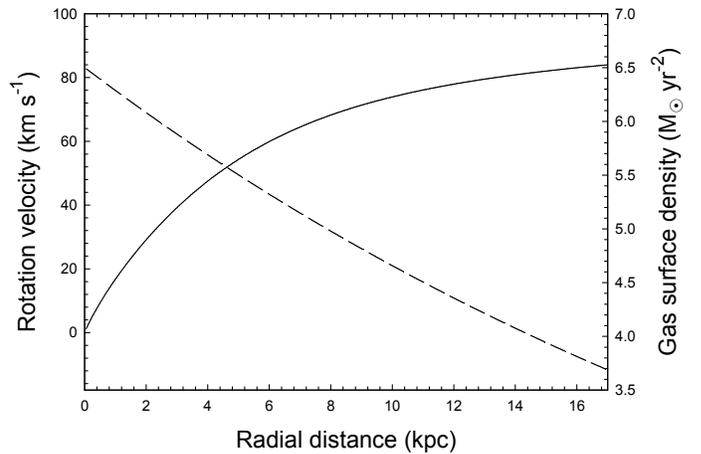}}
      \caption{The initial rotation curve (solid line) and gas surface
      density distribution (dashed line).}
         \label{fig2}
\end{figure}
The measurements of \citet{Kuzio} indicate
that the gas velocity dispersions of LSB galaxies lie in the range between 6~km~s$^{-1}$ 
and 10 km~s$^{-1}$ and are consistent with the typical dispersions for the gas
component of HSB galaxies. Therefore, we assume that the gas disk of our model galaxy 
is {\rm initially} isothermal at a temperature $T=10^{4}$~K. 
LSB galaxies have same atomic hydrogen content as their HSB counterparts
but generally lower surface densities and extended gas disks \citep{Pickering,deBlok96}. 
We assume that the gas disk has an exponentially declining density profile 
\begin{equation}
\Sigma_{\rm g}=\Sigma_{\rm g0} \exp(-r/r_{\rm g}),
\end{equation}
with the central surface density $\Sigma_{\rm g0}=6.5~M_\odot$~pc$^{-2}$
and radial scale length $r_{\rm g}=30$~kpc. 
The gas disk extends to 17~kpc in our model, but star formation is confined to the inner 15~kpc.
The outermost 2~kpc are kept for numerical reasons. 
Gas disks of such extent ($\sim 15$~kpc) are typical for LSB galaxies \citep{Hoek}. 
The mean surface density of our model disk is $\langle \Sigma_{\rm g} \rangle=4.9~M_\odot$~pc$^{-2}$,
which is comparable to a median HI surface density $\Sigma_{\rm HI}=3.8~M_\odot$~pc$^{-2}$ 
of a sample of LSB galaxies in \citet{Hoek}.
The  mass of our model gas disk
in the inner 15~kpc is $M_{\rm g}=3.3 \times 10^{9}~M_\odot$. Hence, the gas disk contains only
a small fraction ($\sim 15\%$) of the total mass in the computational domain
and, to a first approximation, we can neglect its self-gravity.

\begin{table}
\begin{center}
\caption{Main structural properties of our model galaxy
\label{table1}}
\begin{tabular}{clcc}
\hline\hline
& mass & scale length & central density   \\
\hline
 Stellar disk  & $2.6\times 10^9$ & 4  & 30  \\
 gas disk   & $3.3\times 10^9$  & 30  & 6.5  \\
 halo & $2.0\times 10^{10}$  & 5.7  & $6.0\times 10^{-3}$ \\
 \hline
\end{tabular} \\
{All masses are in $M_\odot$, scale lengths in kpc, and  densities in $M_\odot$~pc$^{-2}$ (gas and stellar
disks) and $M_\odot$~pc$^{-3}$ (halo). All masses are calculated inside 15~kpc radius.}
\end{center}
\end{table} 

The initial rotation curve of the gas disk is calculated by solving the steady-state
momentum equation for the tangential component of the gas velocity
\begin{equation}
{u_{\rm \phi}^2 \over r} = {R T \over \Sigma_{\rm g} \mu} {\partial \Sigma_{\rm g} \over \partial r}
+ {\partial \Phi_{\rm st} \over \partial r} + {\partial \Phi_{\rm h} \over \partial r},
\end{equation}
where $\mu=1.2$ is the mean molecular weight for a gas of atomic hydrogen (76\% by mass) 
and atomic helium (24\% by mass) and $R$ is the universal gas constant.
The initial radial gas density profile and rotation curve are shown
in Fig.~\ref{fig2} by the dashed and solid lines, respectively. It is evident that
our model LSB galaxy is characterized by a rising rotation curve with a
maximum rotation velocity of approximately 80~km~s$^{-1}$, in agreement 
with the measured shapes of the rotation curve in UGC~5750 and other LSB
galaxies \cite{Kuzio}. The main properties of our model galaxy are summarized 
in Table~\ref{table1}.

\subsection{Basic equations}
We use the thin-disk approximation to compute the long-term evolution of our model LSB galaxy. 
It is only the gas disk that is dynamically active, while both the stellar disk and dark matter 
halo serve as a source of external gravitational potential.
In the thin-disk approximation, the vertical motions in the gas disk are neglected
and the basic equations of hydrodynamics are integrated in the $z$-direction
from $z=-Z(r,\phi,t)$ to $z=+Z(r,\phi,t)$ to yield the following equations in polar coordinates ($r,\phi$)
\begin{equation}
{\partial \Sigma_{\rm g} \over \partial t} + {{\bl \nabla}_p} \cdot ({\bl v}_p \Sigma_{\rm g}) =
-\beta_{RM} \, \Sigma_{\rm SFR}, 
\label{first}
\end{equation}
\begin{eqnarray}
{\partial {\bl s}_p \over \partial t}  &+& {\bl \nabla_p} \cdot ( {\bl s}_p \cdot {\bl v_p})
= -{\bl \nabla}_p \bar{p} - \Sigma_{\rm g} {\bl \nabla_p} \Phi_{\rm h} \nonumber \\ 
&-&  \Sigma_{\rm g} {\bl \nabla_p} (\Phi_{\rm sp} + \Phi_{\rm st}) - 
{\bl v_p} \beta_{RM}\, \Sigma_{\rm SFR}, 
\end{eqnarray}
\begin{eqnarray}
{\partial \epsilon \over \partial t} + {\bl \nabla_p} \cdot (v_p \epsilon) &=&
-\bar{p} ({\bl \nabla \cdot  v}) + \Gamma_{\rm sn} + 2 Z (\Gamma_{\rm cr}+\Gamma_{\rm bg})
\nonumber \\ &-& 2 Z \Lambda - v_p^2 \beta_{RM}\, \Sigma_{\rm SFR}.
\label{third}
\end{eqnarray} 
In the above equations, $\bl v_p=\hat{r}v_r+\hat{\phi}v_\phi$ is the gas velocity in the 
disk plane, ${\bl s}_p\equiv\Sigma_{\rm g} {\bl v}_{p}$ is the momentum per surface area in the
disk plane,  ${\bl \nabla_p}=\hat{r}\partial
/\partial r + \hat{\phi} r^{-1}\partial /\partial \phi$ is the gradient
along the planar coordinates of the disk, $\bar p$ is the vertically integrated gas pressure,
$\epsilon$ is the internal energy per surface area,
and $Z$ is the vertical scale height of the gas disk.
The basic equations of hydrodynamics are modified to include the effect of star formation,
where  $\Sigma_{\rm SFR}$ is the star formation rate (SFR) per unit area and
$\beta_{\rm RM}=0.42$ is the fraction of stellar mass that 
ends up as an inert remnant \citep{KA}. Equations~(\ref{first})-(\ref{third}) 
are closed with the equation of state of a perfect gas ${\bar p}=(\gamma-1)\epsilon$ and $\gamma=5/3$.

The volume cooling rate $\Lambda\equiv\Lambda\left({\rm erg~cm}^{-3}~{\rm s}^{-1} \right)$
as a function of gas temperature $T$
is calculated from the cooling curves of \citet{Wada} and is linearly interpolated for 
metallicities between $10^{-4}$ and 1.0 of the solar. 
The heating of gas is provided by supernova energy release $\Gamma_{\rm sn}$, cosmic rays
$\Gamma_{\rm cr}$, and background ultraviolet radiation field $\Gamma_{bg}$. 
The supernova energy release (per unit time and unit surface area) 
is defined as
\begin{equation}
\Gamma_{\rm sn}={e_{\rm sn} \Sigma_{\rm SFR} \over m_{\rm sn}}  { \int_{8 M_\odot}^{m_{\rm up}}
 m \, \xi(m) dm  \over 
\int_{m_{\rm low}}^{\rm m_{\rm up}} m \, \xi(m) dm },
\end{equation}
$e_{\rm sn}=10^{51}$~erg is the energy release by a single supernova, and $m_{\rm sn}=15~M_\odot$
is the mean mass of massive stars capable of producing supernova explosions.
We use the Kroupa initial mass function \citep{Kroupa} defined as
\begin{equation}
 \xi(m)= \left\{ \begin{array}{ll} 
   A \, m^{-1.3} & \,\, 
   \mbox{if $m \le 0.5~M_\odot$ } \\ 
   B \, m^{-2.3} & \,\, 
   \mbox{if $m > 0.5~M_\odot$ },  \end{array} 
   \right. 
   \label{function}  
 \end{equation}
with the lower and upper cutoff masses $m_{\rm low}=0.1~M_\odot$ and $m_{\rm up}=100~M_\odot$,
respectively. The ratio of the normalization constants A/B = 2 is found using the continuity 
condition at $m = 0.5~M_\odot$.
We adopt the cosmic ray heating per unit time 
\begin{equation}
\Gamma_{\rm cr} = 10^{-27} n_{\rm g}  \,\, \left( {\rm erg~cm}^{-3}~{\rm s}^{-1} \right),
\end{equation}
where $n_{\rm g}=\Sigma_{\rm g}/(2 Z \mu m_{\rm H})$
is the volume number density of gas. 
Heating due to the background ultraviolet radiation field is supposed to balance cooling
and help sustain a gas temperature $T=10^4$~K in the unperturbed by star formation gas disk of our model
galaxy. Therefore, $\Gamma_{\rm bg}$ is determined in the beginning of numerical simulations 
from the following relation $\Gamma_{\rm bg} n_{\rm g}=\Lambda$, where $\Lambda$ and $n_{\rm g}$
are those of the initial gas disk configuration, and is kept fixed afterwards.


The volume heating and cooling rates in equation~(\ref{third}) are assumed to be independent of $z$-direction
and are multiplied by the disk thickness ($2 Z$) to yield the corresponding vertically integrated rates.
The vertical scale height $Z\equiv Z(r,\phi,t)$ is calculated via the equation of local pressure balance
in the gravitational field of the dark matter halo, stellar, and gas disks
\begin{equation}
\rho_{\rm g} \, c_s^2  = 2\int_0^Z \rho_{\rm g} \left( g_{z}^{\rm gas}+g_{z}^{\rm st} +
g_{\rm z}^{\rm halo} \right) dz,
\label{eq1}
\end{equation}
where $\rho_{\rm g} \, c_s^2$ is the gas pressure in the disk midplane, 
$g_{z}^{\rm gas}$, $g_{z}^{\rm st}$, and 
$g_{z}^{\rm halo}$ are the {\rm vertical} gravitational accelerations due to self-gravity 
of a gas layer, stellar layer, 
and gravitational pull of the dark matter halo, respectively, and $c_s^2=\gamma(\gamma-1)\epsilon/\Sigma_{\rm
g}$ is the square of the sound speed. 
The details of this method are given 
in the Appendix. We plot $Z$ as a function of radius by the dashed line in Fig.~\ref{fig1} 
for our model gas disk in the
beginning of numerical simulations. It is evident that the disk is rather thick, which is indeed expected
for an intermediate-mass galaxy ($\sim {\rm a~few} \times 10^{10}~M_\odot$) considered in the 
present paper.
In practice, we do not let the vertical scale height to drop below 100~pc, since such low values of
$Z$ are not expected in LSB galaxies and may lead to overcooling of our model gas disk.
It should be noticed that in our numerical simulations the energy release by supernova explosions 
is localized in the disk due to the adopted thin-disk approximation. In reality, some portion 
of the supernova energy may be lost to the intergalactic medium if the vertical scale height of a
 gas disk is sufficiently small and the energy release by multiple supernova explosions is 
sufficiently large. Given that intermediate-mass LSB galaxies are characterized by thick 
gas disks and low rates of star formation ($\sim 0.1~M_\odot$~yr$^{-1}$), we do not expect 
this blowout effect to considerably influence the dynamics of our model gas disk.

\subsection{Instantaneous recycling approximation}
\label{hydro}
The temporal and spatial evolution of oxygen  in our model galaxy is computed adopting 
the so-called instantaneous recycling approximation, in which an instantaneous enrichment
by freshly synthesized elements is assumed. This approximation
is valid for the oxygen enrichment, because most of oxygen 
is produced by short-lived, massive stars. We adopt the oxygen yields of 
\citet{WW}. We assume that oxygen is collisionally coupled to the gas, which eliminates the need
to solve an additional momentum equation for oxygen.  The continuity equation for 
the oxygen surface density $\Sigma_{\rm ox}$
modified for oxygen production via star formation reads as

\begin{equation}
{\partial \Sigma_{\rm ox} \over \partial t} + {\bl \nabla_p} \cdot ({\bl v_p} \Sigma_{\rm ox})
=R_{\rm ox} -\beta_{\rm RM} \, \Sigma_{\rm SFR} {\Sigma_{\rm ox} \over \Sigma_{\rm g} }.
\label{oxygen}
\end{equation}
The oxygen production rate per unit area by supernovae is calculated as
\begin{equation}
R_{\rm ox}= \Sigma_{\rm SFR} { \int_{12 M_\odot}^{m_{\rm up}} p(m)\, \xi(m) dm  \over 
\int_{m_{\rm low}}^{\rm m_{\rm up}} m \, \xi(m) dm },
\end{equation}
where $p(m)$ is the mass of oxygen released by a star with mass $m$ \citep{WW}. 
We assume that the initial oxygen abundance of our model galaxy is ${\rm [O/H]}=\log_{10}(\Sigma_{\rm ox}/\Sigma_{\rm g})- \log_{10} \left(\Sigma_{\rm ox}/\Sigma_{\rm g}\right)_\odot =-4$, where 
the ratio $\left(\Sigma_{\rm ox}/\Sigma_{\rm g}\right)_\odot$  in the Solar neighbourhood is set
equal to $7.56\times 10^{-3}$.

\subsection{Code description}
\label{code}
An Eulerian finite-difference code is used to solve equations~(\ref{first})-(\ref{third}) and (\ref{oxygen})
in polar coordinates ($r,\phi$). The basic algorithm of the code is similar
to that of the commonly used ZEUS code. The operator
splitting is utilized to advance in time the dependent variables in two
coordinate directions. The advection is treated using the consistent
transport method and the van Leer interpolation scheme. 
There  are a few modifications to the usual ZEUS methodology, which are necessitated by the presence
of star formation and cooling and heating processes in our model hydrodynamic equations.
The total hydrodynamic time step 
is calculated as a harmonic average of the usual time step due to the Courant-Friedrichs-Lewy condition
and the star formation time step defined as $t_{\rm SF}= 0.1 \Sigma_{\rm g}/\Sigma_{\rm SFR}$.
The internal energy update due to cooling and heating in equation~(\ref{third})
is done implicitly via Newton-Raphson iterations, supplemented by a bisection
algorithm for occasional zones where the Newton-Raphson
method does not converge. This ensures a considerable economy in the CPU time.
A typical run takes about four weeks on four 2.2~GHz Opteron processors.
The resolution is $500\times 500$ grid zones, which are equidistantly spaced in both 
coordinate directions. The radial size of the grid cell is 34~pc and the radial extent of the computational
area is 17~kpc. However, star formation is confined to only the inner 15~kpc.
The code performs well on the angular momentum conservation problem
\cite[see e.g.][]{VB}. This test problem is essential for the adequate modelling of
rotating systems with radial mass transport.

\subsection{Sporadic star formation}
\label{starform}

Low gas surface densities \citep{deBlok96,Pickering}, which are usually below the
critical value determined by the Toomre criterion, and low total star formation rates
\citep{Hoek} argue against
large-scale star formation as a viable scenario for LSB galaxies. 
Despite the low gas surface densities, there is evidence for a noticeable 
young stellar population as indicated by the blue colors of most LSBs 
\citep{deBlok95}. According to \citet{deBlok95} and \citet{Gerretsen}, 
a sporadic star formation scenario (i.e. small surges in the SFR, either 
superimposed on a very low constant SFR or not)
can best explain the observed blue colors and evidence for young stars.

There is little known about physical processes that control sporadic star
formation in LSB galaxies. The recent uncalibrated  H$\alpha$ imaging
of a sample of LSB galaxies  by \citet{Auld} and the observational data presented here 
(see Sect.~\ref{observe}) suggest that star formation is highly clustered.
Current star formation is localized to a handful of
compact regions. There is little or no diffuse H$\alpha$ emission coming
from the rest of the galactic disk.
Motivated by this observational evidence, we adopt the sporadic scenario for star formation
in our model LSB galaxy.

We use a simple model of sporadic star formation in which the individual star formation
sites (SFSs) are distributed randomly throughout the galactic disk and the star formation
rate per unit area in each SFS is determined by a Schmidt law
\begin{equation}
{\Sigma_{\rm SFR} \over {\rm M_\odot~yr^{-1}~kpc^{-2}}} =
\alpha_{\rm SF}~\left( \Sigma_{\rm g} \over 
{\rm M_\odot~pc^{-2}} \right)^{1.5} \,\, {\rm if}~T<T_{\rm cr}, 
\label{Schmidt}
\end{equation}
where $\alpha_{\rm SF}=6\times 10^{-4}$ is a proportionality constant and $T_{\rm cr}=2 
\times 10^4$~K. If the gas temperature in a randomly chosen SFS exceeds the critical value, the
SFS is rejected. The adopted value of $T_{\rm cr}=2\times 10^4$~K differs from typical temperatures
of star-forming molecular clouds: $T \la 30-50$~K. They are however more appropriate for identification
of star-forming regions on $\sim 34$~pc scales that are resolved in the present numerical simulations.
A Schmidt law with index $n\sim 1.5$ would
be expected for self-gravitating disks, if the SFR is equal to the ratio
of the local gas volume density ($\rho$) to the free-fall time ($\propto
\rho ^{-0.5}$), all multiplied by some star formation efficiency $\epsilon$.
The efficiency $\epsilon$ is a measure of the fraction of the gas mass
converted into stars before the clouds are disrupted.
It can be shown that $\epsilon$ is related to $\alpha_{\rm SF}$ as \citep{Vorobyov03}
\begin{equation}
\alpha_{\rm SF}=0.12 \, \epsilon \, \left( {Z \over {\rm pc}}\right)^{-0.5} 
\end{equation} 
This means that the adopted star formation efficiency in our model ranges between
0.045 and 0.15 for possible values of the vertical scale height $Z=100-1500$~pc.

The number of star formation sites at any given time can be roughly determined using the uncalibrated
H$\alpha$ images of LSB galaxies by \citep{Auld}. 
These images  highlight only a handful of small regions that 
have recently formed stars. 
There is little diffuse $H\alpha$ emission, contrary to the emission in K-, R-, and B-bands
which (as a rule) show a noticeable diffuse component. 
The number of HII regions within an individual LSB galaxy identified by \citet{Hoek} ranges between
4 and 26, with a median number of HII regions equal to 11.
Therefore, the number of SFS, $N_{\rm SFS}$, in our numerical simulations is set to approximately 15.
If we further assume that both $N_{\rm SFS}$ and a typical area occupied by an individual 
star formation site  $S_{\rm
SFS}$ do not
change significantly with time, this would yield a near-constant mean SFR
over the lifetime of our model galaxy. 
However, there is evidence that star formation in blue LSB galaxies
does not proceed at a near-constant rate. Our own numerical simulations and modeling by \citet{zack06}
indicate that the SFR should be declining with time to reproduce the
observed H$\alpha$ equivalent widths in LSB galaxies. Therefore, we assume that the typical
area occupied by a SFS is a decreasing function of time of the form $S_{\rm SFS}=S_{\rm SFS0}\exp(-t/\tau_{\rm
SFR})$, where $S_{\rm SFS0}=450\times450$~pc is the area at $t=0$ and $\tau_{\rm SFR}=1.3$~Gyr 
is a characteristic time of exponential decay in the star formation rate.


The duration of star formation in an individual star formation site ($\tau_{\rm
SFS}$) is poorly known. We use the theoretical and empirical knowledge
gained by analyzing similar processes in the Solar neighbourhood. 
Observations of the current star formation activity in 
nearby molecular cloud complexes indicate that the ages of T Tauri stars usually fall in the
$1-10$~Myr range \citep{Palla}.  Clusters that are 10~Myr old often have little associated
gas remaining, implying that the process shuts off by then. However, the volume gas density
in disks of LSB galaxies is expected to be lower (on average) than that of the Solar neighbourhood,
which may prolong the process of gas conversion into stars. It should also be noted
that individual star formation sites in LSB galaxies have characteristic sizes
that are much {\rm larger} than those of an individual giant molecular cloud. Otherwise,
we would have to assume unrealistically high star formation efficiency in LSB galaxies
to retrieve the observed star formation rates of order $0.1~M_\odot$~yr$^{-1}$.
This implies a presence of {\it local} triggering mechanisms within an individual SFS 
that should also act to prolong the star formation activity.
From a theoretical background one can argue that the duration of star formation 
should be limited by a feedback action from
newly born massive stars and supernova explosions.  The lifetime
of a least massive star capable of producing a supernova is approximately
40~Myr.  Hence, the actual lifetime of an individual SFS should lie in between 10~Myr and 40~Myr.
Using spectro-photometric evolution models, \citet{Hoek} have also found that $\tau_{\rm SFS}=10-50$~Myr
reproduce well the observed $(R-I)$ colors and $I$-band luminosities of LSB galaxies.
In our numerical simulations we assume that $\tau_{\rm SFS}=20$~Myr. Thus, each twenty 
million years we initialize a new set of SFSs ($N_{\rm SFS}\approx 15$) 
and randomly distribute them in the inner 15~kpc of the galactic disk. 
Each SFS is assigned a velocity equal to that of the local gas velocity upon
its creation and is evolved collisionlessly in the total gravitational 
potential of the stellar disk and the dark matter halo using the
fifth-order Runge-Kutta Cash-Karp integration scheme. The old SFSs are rendered inactive
and their further evolution is not computed.

\section{Observations}
\label{observe}
\begin{figure}
  \resizebox{\hsize}{!}{\includegraphics[angle=-90]{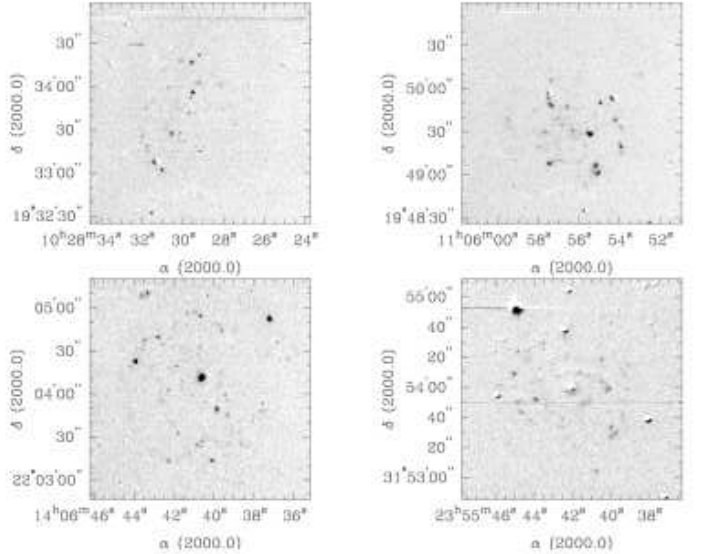}}
      \caption{Reprocessed  and continuum subtracted images of the 
LSB galaxies 
UGC 9024 (upper left), UGC 12845 (upper right), 
UGC 5675 (lower left), and UGC 6151 (lower right).}
         \label{fig3}
\end{figure}

There is still just a small number of (nearby) LSB galaxies 
with published H$\alpha$ images, which would allow the study 
of the distribution and properties of their star-formation regions. 
H$\alpha$ images of about 18 LSB galaxies are available in 
\citet{McGaugh1994,McGaugh1995}, 
while \citet{deBlok98} only published the spectra and 
some broad band images. A new sample of 6 LSB galaxies was 
presented in \citet{Auld}.   

From the image data set (McGaugh, priv. comm.) 
we selected 4 LSB spiral galaxies based on their B band morphology.
We selected the galaxies to be close to face-on and showing different 
disk surface brightness and spiral morphology. This resulted in 
UGC 9024, UGC 12845, UGC 5675, and UGC 6151. 
A different version of the H$\alpha$ images of UGC 9024, UGC 5675, 
and UGC 6151 was previously published in \citet{McGaugh1995}.

For the presentation in Figure \ref{fig3} we improved on the cosmic rays 
rejection and correction for detection blemishes of the 
already flat-fielded images. The generation of the continuum corrected 
H$\alpha$ images followed closely the schema described 
e.g. in \citet{Bomans1997}.
We realigned the H$\alpha$ and continuum 
images carefully using the foreground stars visible in 
both images to sub-pixel accuracy and subtracted the scaled 
continuum image. We tested different scaling factors and 
defined an optimal subtraction based on the disappearance of the 
stellar disk of the LSB galaxy.  These factors also gave the 
smallest foreground star residuals and is consistent with 
the scaling factor calculated by measuring the flux of several 
foreground stars in both images. 

The distribution of the HII regions in all 4 galaxies in Figure \ref{fig3}  
is consistent with the predicted pattern of sporadic star formation.  
The H$\alpha$ images of 6 LSB galaxies recently published by 
\cite{Auld} show the same behavior, as do the other 
galaxies published in \cite{McGaugh1995}.

\section{Numerical results}
\label{results}
\subsection{Gas density and temperature distribution}

\begin{figure}
  \resizebox{\hsize}{!}{\includegraphics{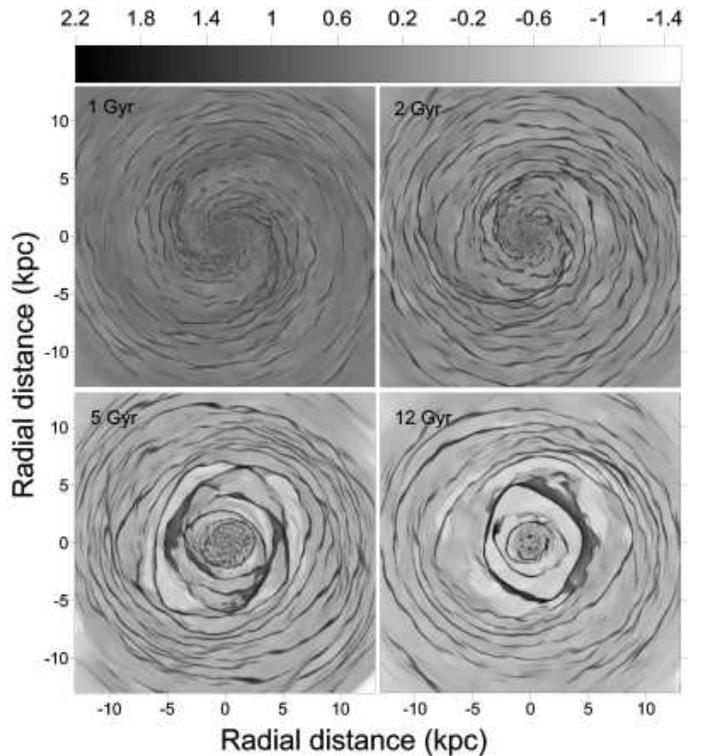}}
      \caption{Gas surface density distribution in the 1-Gyr-old (upper left),
      2-Gyr-old (upper right), 5-Gyr-old (lower left), and 12-Gyr-old (lower right) disks.
      Dark regions are the densest and coldest. The scale bar is $M_\odot$~pc$^{-2}$ (log units).}
         \label{fig4}
\end{figure}

A model galaxy described in Section~\ref{modelgalaxy} will be 
referred hereafter as model~1. In a due course, we will also introduce other
models that will have identical structural properties such as those summarized in Table~\ref{table1}
but different IMFs and star formation rates.
We start our numerical simulations by setting the equilibrium gas disk and slowly introducing 
the non-axisymmetric gravitational potential of the stellar spiral arms. Specifically, 
$\Phi_{\rm sp}$ is multiplied by
a function $\xi(t)$, which has a value of $0$ at $t=0$ and linearly
grows to its maximum value of $1.0$ at $t\ge200$~Myr.
It takes another hundred million years for the gas disk to adjust to the spiral gravitational potential
and develop a spiral structure. Figure~\ref{fig4} shows the gas 
surface density distribution (in log units) in the 1-Gyr-old (upper left), 
2-Gyr-old (upper right), 5-Gyr-old, (lower left), and 12-Gyr-old (lower right) disks.
It is seen that the gas disk develops a strong phase separation with age.
While the initial gas disk has surface densities between $3.5~M_\odot$~pc$^{-2}$ and $6.5~M_\odot$~pc$^{-2}$,
the 1-Gyr-old disk features some dense clumps with surface densities in excess of $100~M_\odot$~pc$^{-2}$
and temperatures of order 100--200~K. The number of such clumps increases with time and they are getting
stretched
into extended filaments and arcs by galactic differential rotation. Observational detection
of such features could be a powerful test on the sporadic nature of star formation in LSB galaxies.
The gas response to the underlying spiral stellar density wave is rather weak in the early 
disk evolution and can hardly be noticed in the 5- and 13-Gyr-old disks. It implies that mixing 
due to spiral stellar density waves is inefficient in the old disk.

\begin{figure}
  \resizebox{\hsize}{!}{\includegraphics{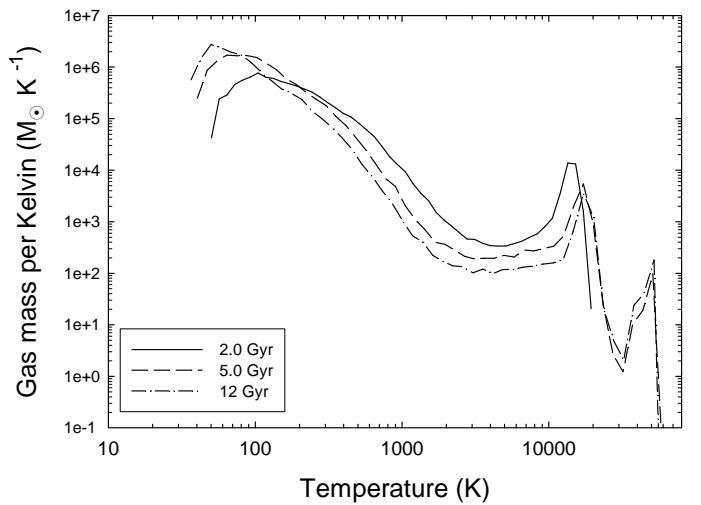}}
      \caption{Gas mass per Kelvin versus temperature in disks
      of distinct age as indicated in the legend. The two peaks manifest the phase separation.
      Note that there is little gas with temperatures below 40~K. }
         \label{fig5}
\end{figure}

The phase separation of the gas disk can be nicely illustrated
by calculating the gas mass per unit temperature band at different evolutionary times. 
Figure~\ref{fig5} shows the gas mass per Kelvin as a function of temperature at $t=2$~Gyr (solid line),
$t=5$~Gyr (dashed line) and $t=13$~Gyr (dash-dotted line). There are two prominent peaks that
manifest the separation of the gas into two phases -- a cold gas phase with peak temperature about
$T=50-100$~K and a warm gas phase with peak temperature about $T=(1.0-1.5)\times 10^4$~K. 
The positions of the peaks drift apart as the gas disk evolves, indicating that the cold phase becomes
colder and the warm phase becomes warmer with time. However, the gas temperature never drops below 35~K.
It is also evident from the amplitudes of the peaks that the cold phase accumulates 
gas mass with time, which is explained by increased cooling due to ongoing heavy element 
production. A similar phase separation was also obtained in numerical simulations of LSBs by \citet{Gerretsen}.
The fact that there is little gas with temperature below 35~K implies
that most of the cold phase may be in the form of atomic rather than molecular hydrogen,
but more accurate numerical simulations with a chemical reaction network and higher
numerical resolution are needed to prove this hypothesis.


Figure~\ref{fig5} indicates that model~1 has no hot gas phase with temperatures of
order $10^6$~K, which is a consequence of a relatively low SFR and moderate numerical resolution. 
The energy injection by star formation of such a low rate
is not enough to convert all gas in a computational cell ($\sim 34 \times 34$~pc) to a hot phase. 
To remind the reader, we solve the equations of hydrodynamics for a single fluid. 
As a consequence, temperature of {\it all} gas
in a computational cell is characterized by a specific value, 
there is no gas 
phase separation  within an individual cell. This, however, does not mean that the hot gas 
is not present at all 
in LSB galaxies, we simply may need a much higher numerical resolution to resolve it.
We do start seeing the hot phase if we increase the SFR in our model by a factor of several.
We note that \citet{Gerretsen}
have also found little hot gas in their numerical simulations.

\begin{figure}
  \resizebox{\hsize}{!}{\includegraphics{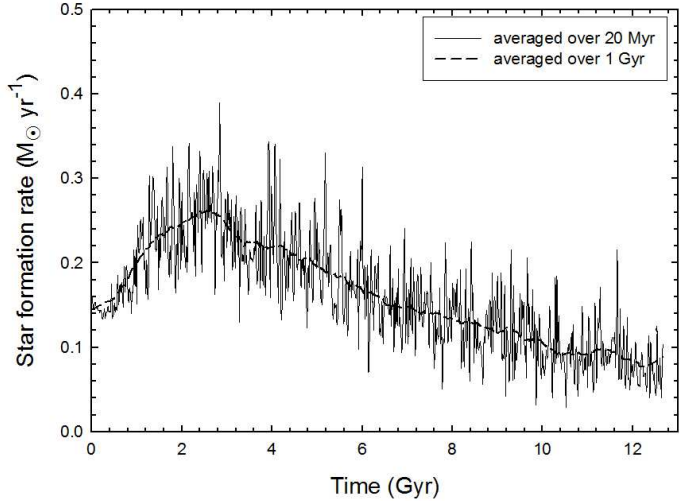}}
      \caption{Star formation rate versus time in model~1. The solid and thick dashed 
      lines show the 
      SFR averaged over 20~Myr and 1.0~Gyr, respectively. Note large amplitude 
      fluctuations in the 20-Myr-averaged SFRs around the 1-Gyr-averaged values.}
         \label{fig6}
\end{figure}

Figure~\ref{fig6} shows the SFR as a function of time in model~1.
In particular, the solid line presents the SFR averaged over the past 20~Myr, these values
are later used to construct H$\alpha$ equivalent widths. 
In addition, the thick dashed line plots the SFR averaged over a 1~Gyr interval centered
on the current time (running average method).
It is seen that the 20-Myr-averaged SFRs have considerable fluctuations around 
the 1-Gyr-averaged values.
These fluctuations have a characteristic time scale of a few hundreds of Myr and 
amplitudes that may exceed $0.1~M_\odot$~yr$^{-1}$. 
The existence of such bursts of star formation is an essential ingredient for successful modelling
of blue colors of at least some LSB galaxies \citep[e.g.][]{Hoek}. Similar fluctuations in the SFR 
were also observed in numerical simulations by \citet{Gerretsen}. 
The 1-Gyr-averaged SFR quickly grows to a maximum value $0.26~M_\odot$~yr$^{-1}$ 
at $t=2.6$~Gyr and subsequently declines to $0.08~M_\odot$~yr$^{-1}$ at $t=13.0$~Gyr.
Typical star formation rates in LSB galaxies range 
from 0.02 to $0.2~M_\odot$~yr$^{-1}$, with a median SFR of $0.08~M_\odot$~yr$^{-1}$ \citep{Hoek}. 
Model~1 yields SFRs that are consistent with the inferred present-day rates.

\subsection{Radial variations in the oxygen abundance}
We calculate the oxygen abundance ${\rm [O/H]}=\log_{10}(\Sigma_{\rm ox}/\Sigma_{\rm g})- 
\log_{10}(7.56\times 10^{-3})$ in our disk from the model's known surface densities of gas 
($\Sigma_{\rm g}$) and oxygen  ($\Sigma_{\rm ox}$).
We find that the spatial distribution of oxygen in the 1-Gyr-old 
gas disk has a pronounced patchy appearance -- local regions with enhanced oxygen abundance
adjoin local regions with low oxygen abundance. The local contrast in [O/H] may be as
large as 1.0~dex or even more. As the galaxy evolves, the spatial
distribution of oxygen becomes noticeably smoother. The smearing is caused by stirring 
due to differential rotation, radial migration of gas triggered 
by the non-axisymmetric gravitational potential of stellar spiral arms, and energy release
of supernova explosions.

\begin{figure}
  \resizebox{\hsize}{!}{\includegraphics{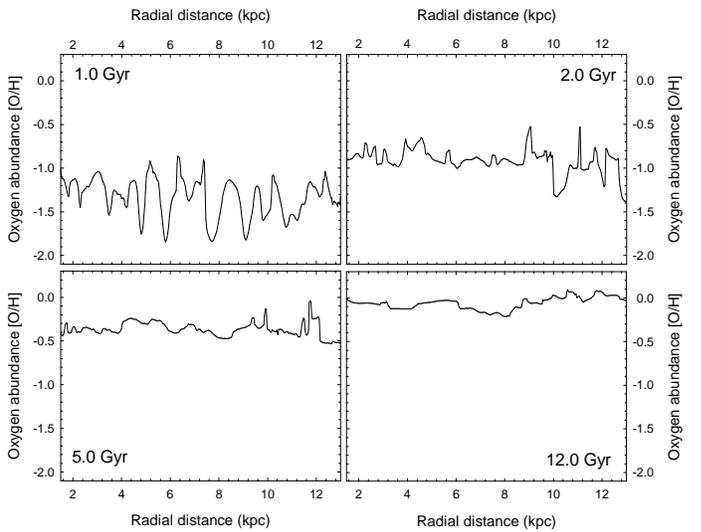}}
      \caption{Radial profiles of [O/H]  obtained by making a narrow cut through
      the galactic disk at an azimuthal angle $\phi=0^\circ$. The age of the disk is 
      indicated in each frame.}
         \label{fig7}
\end{figure}

To better illustrate the spatial variations in the oxygen distribution in model~1, 
we show in Fig.~\ref{fig7} the radial profiles of the oxygen abundance 
obtained by making a narrow ($\sim 50$~pc) radial cut through the galactic disk at 
$\phi=0^\circ$.
The numbers in each frame indicate the age of our model galaxy. The most prominent 
feature in Figure~\ref{fig7} is a sharp decline in a typical amplitude of the radial fluctuations
in [O/H]. For instance, the radial distribution of the oxygen abundance 
in the 1-Gyr-old disk is very irregular and shows radial fluctuations with amplitudes as large 
as 1.0~dex and even more. The fluctuations have a short characteristic radial scale of 
order $1-2$~kpc. The fluctuation amplitudes have noticeably decreased in the 2-Gyr-old disk,
which now has a maximum amplitude of order $0.5$~dex.
The 5-Gyr-old disk (and older disks) feature only low-amplitude fluctuations 
in the oxygen abundance with amplitudes rarely exceeding $0.2$~dex.

\begin{figure}
  \resizebox{\hsize}{!}{\includegraphics{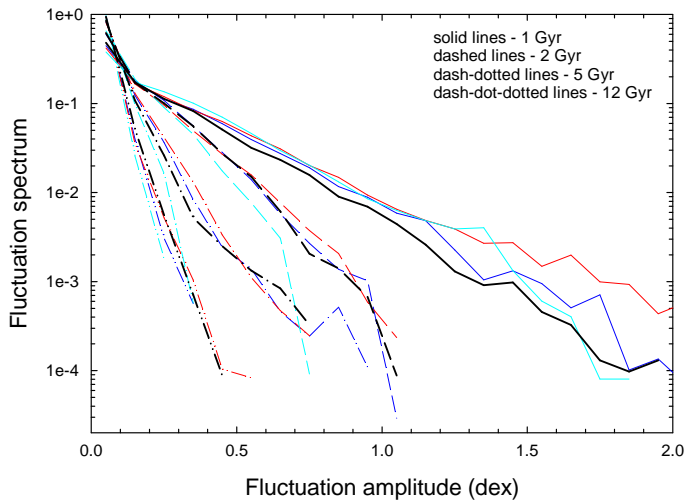}}
      \caption{Oxygen abundance fluctuation spectrum defined as the normalized number of 
      radial oxygen abundance fluctuations versus fluctuation amplitude (in dex).  Lines of
      distinct color corresponds to different numerical
      models and lines of distinct type corresponds to different galactic ages 
      as indicated. See the text for more details.}      
         \label{fig8}
\end{figure}

Figure~\ref{fig7} reveals the absence of any significant radial abundance gradients
in our model disk. This appears to be typical for LSB galaxies \citep[e.g.][]{deBlok98}.
The lack of radial abundance gradients is most likely caused by a spatially and temporally 
sporadic nature of star formation, i.e., LSB galaxies must have evolved at the same rate over their entire
disk. Well-known star formation threshold criteria, such as those based
on the gas surface density threshold (Toomre criterion) or the
rate of shear (\citet{MK,Vorobyov03}), would 
increase the likelihood for star formation in regions of higher density and lower angular velocity,
which are usually the inner galactic regions. As a result, negative radial abundance gradients 
may develop with time. The fact that we do not see radial abundance gradients in LSB galaxies
implies that star formation has no threshold barrier.

To analyze the properties of the oxygen abundance fluctuations 
in disks of different age, we calculate the fluctuation spectrum defined as 
the (normalized) number of fluctuations versus the fluctuation amplitude $A$. 
The latter is 
calculated in the following manner. We scan the disk in the radial 
direction along each azimuthal angle ($\phi$) of our numerical polar grid 
for the adjacent minima and maxima in the oxygen abundance distribution.  
The absolute value of the difference between the adjacent minimum and maximum (or vice versa)
gives us the radial amplitude of a fluctuation on spatial scales of our numerical resolution,
$\sim 40$~pc. The calculated fluctuation amplitudes are distributed in 40 logarithmically spaced
bins between 0.1~dex and 3.0~dex and then normalized by the {\it total} number 
of fluctuations in all bins. The resultant fluctuation spectrum has the meaning 
of the probability function
${\cal F}(A)$---the product of the total number of fluctuations and ${\cal F}(A)$ 
yields the number of fluctuation with amplitude $A$.

Figure~\ref{fig8} shows the probability function
${\cal F}(A)$ derived for four models detailed below. Hereafter, different 
line types in Fig.~\ref{fig8} will help to distinguish between galaxies of distinct age
and different line colors will correspond to different numerical models.
Black lines present ${\cal F(A)}$ in model~1 for the 1-Gyr-old disk (solid line), 
2-Gyr-old disk (dashed line), 5-Gyr-old disk (dash-dotted line), 
and 12-Gyr-old disk (dash-dot-dotted line). It is evident that the probability functions for disks of
different age intermix for ${\cal F}(A)\ga 0.1$ but appear to be quite distinct 
for ${\cal F}(A)\la 0.05$. Can this peculiar feature of ${\cal F}(A)$
be used to constrain the age of LSB galaxies? The extent to which oxygen is spatially mixed 
in the galactic disk may depend on the model parameters such as the strength of the 
spiral density wave, shape of the IMF, and details of the energy release by SNe. 
In order to estimate the dependence of ${\cal F}(A)$
on these factors, we run two test models. In particular, model~T1 has no spiral density waves and
model~T2 is characterized by a factor of 2 lower energy input by SNe (latter `T' stands for `test'
to distinguish from regular models).
Model~T2 attempts 
to mimic the case when part of the SN energy is lost to the extragalactic medium via SN bubbles. 
It is also important to see the effect that a different IMF may have on the probability function.
Therefore, we have also introduced model~2, which is characterized by the Salpeter IMF with
the upper and lower cut-off masses of $0.1~M_\odot$ and $100~M_\odot$. Model~2
has the same structural parameters as model~1 except for $\tau_{\rm SFR}$ which is set to 18.5~Gyr.
The parameters of models~1 and 2 are detailed in Table~\ref{table2}.
The probability functions corresponding 
to model~T1, model~T2, and model~2 are shown in the top panel of Fig.~\ref{fig8} 
by the red, blue, and cyan lines, respectively. The meaning of the line types 
(i.e., solid, dashed, etc.) is indicated in the figure.

A noticeable scatter in the probability 
functions of distinct age is seen
among different models. In particular, the probability functions of galaxies of 5-12~Gyr age  
intermingle and thus cannot be used for making age predictions. However, it is still possible
to differentiate between (1-2)-Gyr-old and (5-12)-Gyr-old galaxies based
on the slope of ${\cal F}(A)$. Young galaxies of 1-2~Gyr age have fluctuation spectra with
a much shallower slope than (5-13)-Gyr-old galaxies, meaning that the former
feature radial oxygen abundance fluctuations of much larger amplitude than the latter.
A similar behaviour of  ${\cal F}(A)$ is found 
when we decrease the spatial resolution of the fluctuation amplitude measurements
from 40~pc to several hundred parsecs (more appropriate for modern observational techniques).
We conclude that measurements of the radial oxygen abundance fluctuations may be used only to 
set {\it order-of-magnitude} constraints on the age of blue LSB galaxies.


\begin{table*}
\begin{center}
\caption{Summary of galactic models}
\label{table2}
\begin{tabular}{lccccc}
\hline
\hline
 model & IMF type & upper/lower cutoffs & $\tau_{\rm SFR}$ & $\alpha_{\rm SF}$ & min/max SFR  \\ 
\hline 
  1 & Kroupa & $0.1-100~M_\odot$ & 13.0 & $6\times 10^{-4}$ & 0.075/0.26  \\ 
  1L & Kroupa & $0.1-100~M_\odot$ & 13.0 & $4\times 10^{-4}$  &  0.085/0.17 \\
  2 & Salpeter & $0.1-100~M_\odot$ & 18.5 & $6\times 10^{-4}$ & 0.045/0.20 \\
  2L & Salpeter & $0.1-100~M_\odot$ & 18.5& $4\times 10^{-4}$ & 0.050/0.14 \\
\hline 
\end{tabular} 

Star formation rate is in $M_\odot$~yr$^{-1}$ and characteristic time
of SFR decay $\tau_{\rm SFR}$ is in Gyr. 
\end{center}
\end{table*}


We are aware of only one successful measurement of the radial abundance 
gradients in three LSB galaxies done by \citet{deBlok98}. Unfortunately, the number 
of resolved HII regions and oxygen abundance measurements in each galaxy is too low ($< 10$) 
to make a meaningful comparison with our predictions.
With such a small sample, the probability function would be confined in the $0.1<{\cal F}(A)<1.0$ range,
which is not helpful for making age predictions due to degeneracy of ${\cal F}(A)$ in this range.

\section{Constraining the age of blue LSB galaxies}
\label{synthesis}
In this section, we investigate if the time evolution of the 
mean oxygen abundance, 
complemented by H$\alpha$ emission equivalent widths
and optical colors, can be used to better constrain the age of blue LSB galaxies. 
We consider several new models (in addition to our models~1 and 2) that have different 
rates of star formation.

\subsection{Time evolution of the oxygen abundance}
\label{secoxygen}

\begin{figure}
  \resizebox{\hsize}{!}{\includegraphics{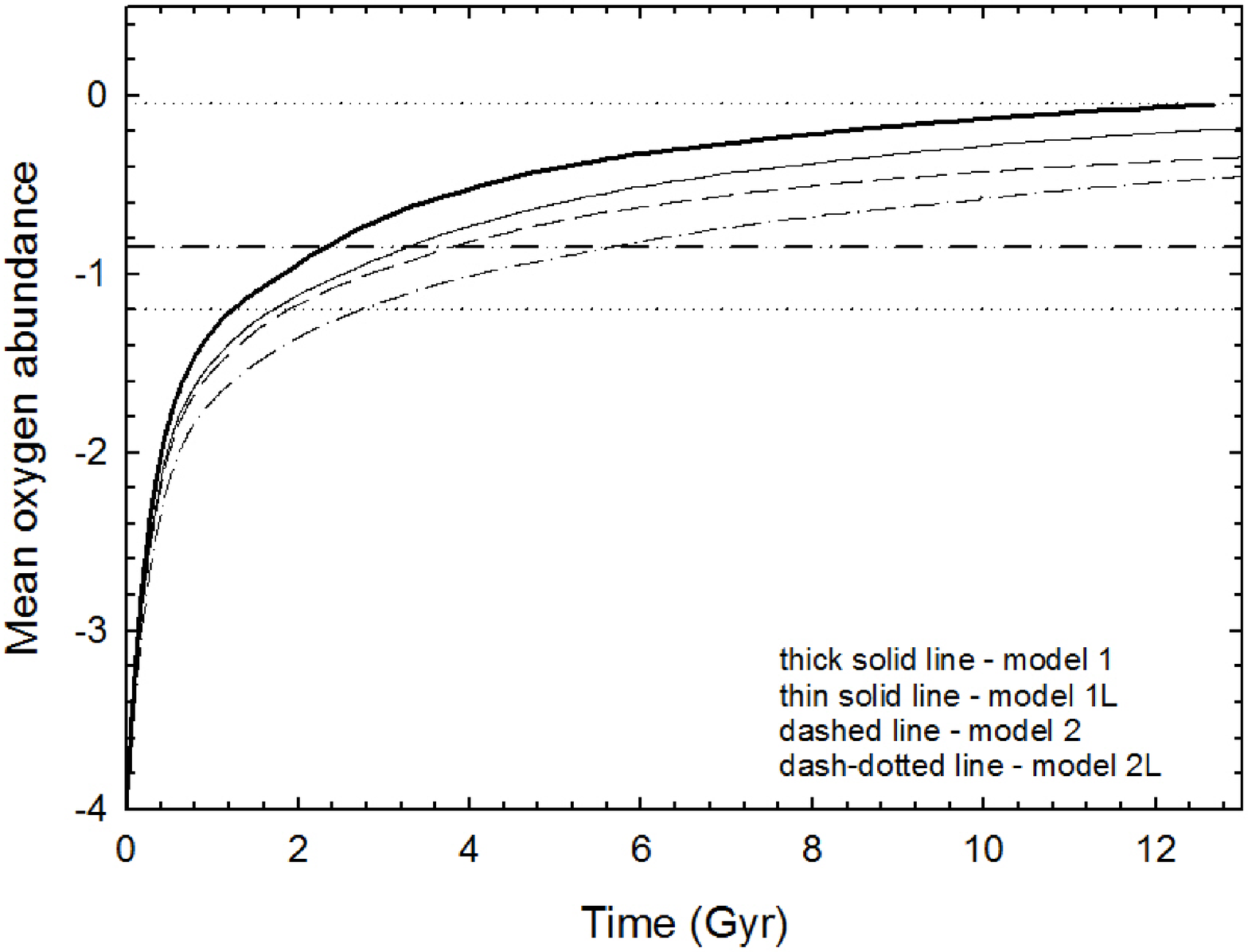}}
      \caption{Mean oxygen abundance as a function of galactic age. 
      Each line type corresponds to a particular numerical model as indicated in the figure. 
      The horizontal dotted lines delineate the range of oxygen abundances found in 
      blue LSB galaxies \citep{deBlok98,Kuzio04} and the horizontal dashed-dot-dotted line shows the
      mean oxygen abundance derived from a large sample of LSB galaxies.
      See the text and Table~\ref{table2} for more details.}
         \label{fig9}
\end{figure}

There have been several efforts in the past to measure the oxygen abundances in HII 
regions of blue LSB galaxies. For instance, oxygen abundances in a sample of 18 
galaxies in \citet{Kuzio04} have a median value $\langle {\rm [O/H]} \rangle = -0.95$. 
Their smallest and largest measured abundances (using the equivalent width method) 
are ${\rm [O/H]}=-1.2$ (F611-1) and ${\rm [O/H]}=-0.04$ (UGC~5709).
\citet{deBlok98} have observed 64 HII regions in 12 LSB galaxies and derived oxygen abundances
that peak around ${\rm [O/H]}=-0.80$. The lowest abundance in their sample is ${\rm [O/H]}=-1.1$.
These data demonstrate that the present day mean oxygen abundance in LSB galaxies is about 
$\langle{\rm [O/H]}\rangle_{\rm LSB}=-0.85$~dex, albeit with a wide scatter. 
We plot the minimum and maximum oxygen abundances found in blue LSB galaxies by horizontal dotted 
lines in Fig.~\ref{fig9}, while the mean oxygen abundance 
for LSB galaxies is shown by the vertical dash-dot-dotted line.  

The amount of produced oxygen and the oxygen abundance are expected to depend 
on the adopted rate of star formation and on the details of the initial mass function.
Most of oxygen is produced by intermediate to massive stars and both the slope and 
the upper mass cutoff are those parameters of the IMF that may influence 
the rate of oxygen production. In this paper, we study IMFs of different slope, leaving 
cut-off IMFs for a subsequent study.


The mean oxygen abundances $\langle {\rm [O/H]}\rangle$ obtained in
model~1 are presented in Fig.~\ref{fig9} by the thick solid line. The mean values are 
calculated by summing up local oxygen abundances in every computational cell
between 1~kpc and 14~kpc and then averaging over the corresponding number of cells.
The dashed line shows  $\langle {\rm [O/H]}\rangle$ as a function 
of galactic age in model~2. In addition, we have introduced two more models that have structural 
parameters identical to those of models~1 and 2 (see Table~\ref{table1}) but are characterized 
by lower SFRs. In particular, models~1L and 2L (thin solid line/dash-dotted line) feature on 
average a factor of 1.4 lower rates of star formation than models~1 and 2.
The parameters of models~1L and 2L, as well as the minimum and maximum SFRs, 
are listed in Table~\ref{table2}. It is evident that models with Salpeter IMF yield 
lower $\langle {\rm [O/H]}\rangle$ than the corresponding models with Kroupa IMF.
With the lower and upper mass limits being identical, the Sapleter IMF may 
be regarded bottom-heavy 
in comparison to that of Kroupa. A bottom-heavy IMF has a lower relative number of massive stars
and this fact accounts for a lower rate of oxygen production in models with the Salpeter IMF.
Furthermore, models with lower SFRs yield lower mean oxygen abundances.

There are several interesting conclusions that can be drawn by analyzing 
Fig.~\ref{fig9}. First, our models can reproduce the entire range of observed oxygen abundances
found in LSB galaxies (delimited by horizontal dotted lines). 
Second, for a given value of $\langle {\rm [O/H]} \rangle$, the predicted ages show a large scatter
(that increases along the sequence of increasing oxygen abundances),
making mean oxygen abundances {\it alone} a poor age indicator, especially for metal-rich LSBs. 
Finally, the minimum oxygen abundance found among LSB galaxies
(bottom dotted line in Fig~\ref{fig9}) can be attained after just
1--3.5~Gyr of evolution, suggesting a possibility of rather young age
for LSB galaxies. We will return to this issue in some more detail 
in the following section.

\begin{figure}
  \resizebox{\hsize}{!}{\includegraphics{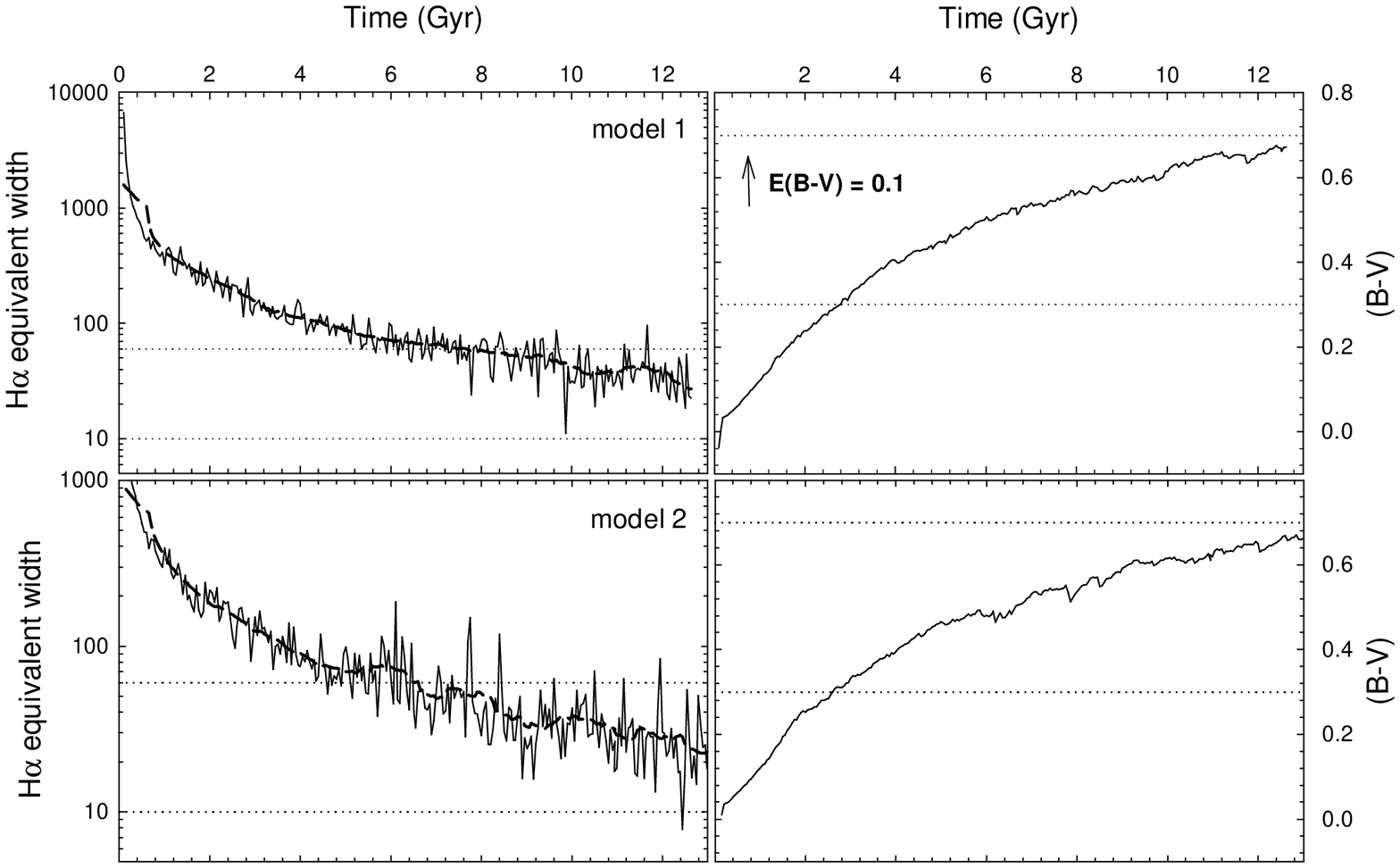}}
      \caption{H$\alpha$ equivalent widths (left column) and global $B-V$ colors (right column) for
      model~1 (top row) and model~2 (bottom row). The horizontal dotted lines delineate a 
      range of observed
      values in blue LSB galaxies according to \citet{deBlok95} and \citet{zack06}. The thick dashed
      lines show $EW$H$\alpha$ averaged over 1~Gyr. The arrow shows a (possible) reddening due 
      to dust extinction for a fiducial value of the color excess $E(B-V)=0.1$ \citep{mcg94}.}
         \label{fig10}
\end{figure}

\subsection{H$\alpha$ equivalent widths and B-V integrated colors}

In this section we use population synthesis modelling to 
calculate the integrated $B-V$ colors and H$\alpha$ equivalent widths ($EW$(H$\alpha$)) using the model's
known SFRs and mean oxygen abundances. The interested reader can find details of the population
synthesis model and essential tests in the Appendix. 
There is observational evidence that LSB galaxies are characterized by $B-V$ colors and $EW$(H$\alpha$)
that lie in certain limits. For instance, the observed $B-V$ colors of blue LSB galaxies 
(which we attempt to model in this paper) appear to lie in the $0.3-0.7$ range \citep{deBlok95}.
According to \citet{zack06}, $EW$(H$\alpha$) of a sample of blue LSB galaxies seem to be confined 
within rather narrow bounds, $EW({\rm H}\alpha)=10-60$~\AA. Hence, population synthesis modelling 
may be regarded as a consistency test for our numerical models, which should yield 
$EW$(H$\alpha$) and $B-V$ colors in acceptable agreement with observations. 

We present the results of population synthesis modelling only for models with different
IMFs; models with different SFRs show a similar behaviour. Figure~\ref{fig10} 
shows model $EW$(H$\alpha)$ (left column) and $B-V$
colors (right column) as a function of galactic age for model~1 (top)
and model~2 (bottom). The equivalent widths are derived using SFRs averaged over the past 20~Myr. 
For convenience, the thick dashed lines show $EW$(H$\alpha$) averaged over 1~Gyr.
The horizontal dotted lines in the left 
panels show a typical range for $EW$(H$\alpha$) in blue LSB galaxies \citep{zack06},
while the horizontal dotted lines in the right panels outline typical $B-V$ colors of blue 
LSB galaxies \citep{deBlok95}. The arrow shows a (possible) reddening due to dust extinction
for a fiducial value of the color excess $E(B-V)=0.1$ \citep{mcg94}.

It is evident that model $EW$(H$\alpha$) show considerable
fluctuations of order 15--20~$\AA$ around mean values, which reflect the corresponding 
fluctuations in the SFR. Nether model~1 nor model~2 can account for the observed equivalent widths
in the early evolution, yielding too large $EW$(H$\alpha$) during the first 5--6~Gyr. 
Both models begin to match the observed $EW$(H$\alpha$) only in the late evolution after 6~Gyr, 
with model~2 (Salpeter IMF) yielding a somewhat better fit.
We note that all models feature short-term variations in $EW$(H$\alpha$)
of such an amplitude that the corresponding equivalent widths may accidently go beyond either the 
upper or lower observed limits in the late evolution phase. 
This is however not alarming, since these fluctuations are short-lived and the
observed range is inferred from
just a few galaxies and may enlarge as more observational data become available.


Unlike model $EW$(H$\alpha$), there are no large short-term fluctuations
in model $B-V$ colors shown in the right column of Fig.~\ref{fig10}. 
Younger galaxies have on average bluer colors. The lowest observed color $B-V=0.3$ 
is reached after approximately 2.5--3.0~Gyr of evolution.
Taking into account a possible reddening, the latter value may decrease to approximately
1.5--2.0~Gyr. A similar lower limit on the galactic age was inferred from
numerical modelling of oxygen abundances in Section~\ref{secoxygen}.
However, model equivalent widths suggest a much larger value, $\sim 5-6$~Gyr.
This value may decrease slightly, if LSB galaxies feature truncated IMFs with a smaller
upper mass limit, say 40~$M_\odot$ instead of 100~$M_\odot$.
The oxygen abundance fluctuation spectrum ${\cal F}(A)$ can be used to better constrain the minimum
age of blue LSB galaxies. 

It is well known that 
blue LSB galaxies have $B-V$ colors that are systematically bluer than those of HSB galaxies
\citep{deBlok95}. It is tempting to assume that this fact may be indicative of blue LSB 
galaxies being systematically younger than their HSB counterparts. However, our
evolved galaxies seem to have colors that fall within the
observed limits. There is a mild possibility for the model $B-V$ color to exceed 
the upper observed limit if dust extinction is taken into account. But considering
the uncertainty in dust content in blue LSB galaxies, this possibility 
is not conclusive. In any way, even if blue LSB galaxies are younger
than their HSB counterparts, the difference in age is not dramatic,
of the order of a few Gyr at most. It is likely that blue LSB galaxies owe their bluer 
than usual colors to the sporadic 
nature of star formation \citep[see e.g.][]{mcg94,deBlok95,Gerretsen}.



\subsection{Mean oxygen abundances versus H$\alpha$ equivalent widths}
\begin{figure}
  \resizebox{\hsize}{!}{\includegraphics{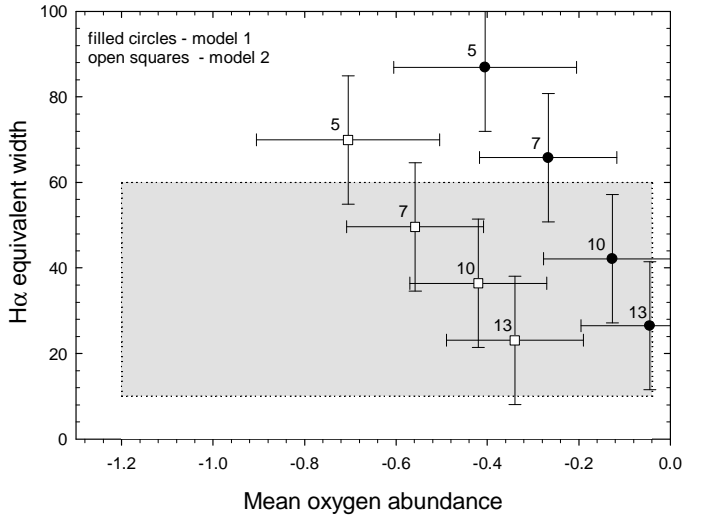}}
      \caption{Mean oxygen abundance versus H$\alpha$ equivalent width diagram. Each symbol 
      presents a pair of data points  
      corresponding to a model galaxy of specific age, as indicated in Gyr. 
      Each symbol type corresponds
      to a particular numerical model specified in the upper-left corner. Bidirectional bars 
      assigned to every symbol characterize a typical range of 
      values found in our numerical modelling. A gray rectangular area
      delineates a range of observed values typical for blue LSB galaxies.  }
         \label{fig11}
\end{figure}

As the final stroke of the brush, we present the $\langle {\rm [O/H]} \rangle - 
EW({\rm H}\alpha)$ diagram in Fig.~\ref{fig11}. Each symbol in this diagram 
represents a pair of data points $[\langle {\rm [O/H]} \rangle, EW({\rm H}\alpha)]$ 
corresponding to a specific galactic age in Gyr as indicated in the figure. In particular, 
filled circles and open squares represent models~1 and 2, respectively.
Each symbol is assigned bi-directional bars, which characterize a typical range of values
found in our numerical simulations. More specifically, the vertical bars illustrate typical
short-term fluctuations of $EW({\rm H}\alpha)$ within each model, while the horizontal bars
represent a typical range of $\langle {\rm [O/H]} \rangle$ found among models with identical IMFs but
different SFRs. 
A gray rectangular area shows a typical range of observed oxygen abundances and 
H$\alpha$ equivalent widths in blue LSB galaxies.

There are several important features in Fig.~\ref{fig11} that we want to emphasize.
First, only those model galaxies that are older than 5--7~Gyr appear to fit into 
the observed range of both $EW({\rm H}\alpha)$ and $B-V$. Younger galaxies fail to
reproduce the observed H$\alpha$ equivalent widths.
Second, numerical models with standard mass 
limits ($m_{\rm low},m_{\rm up}=0.1,100~M_\odot$) populate the upper-right portion 
of the observed $EW({\rm H}\alpha)$ vs. $B-V$ phase space. This suggest that at least some
LSB galaxies may have IMFs that are different from those considered in the
present work, i.e., they may have a truncated upper mass limit.
Third, our modelling suggests an upper limit on the age of blue LSB galaxies that is 
comparable to or slightly less than the Hubble time.
Forth, an accurate age prediction is possible if some independent knowledge of the IMF
is available.

\section{Conclusions} 
\label{conclude}
We have used numerical hydrodynamic modelling to study the long-term 
($\sim 13$~Gyr) dynamical, chemical, and photometric evolution of blue LSB galaxies.
Motivated by observational evidence, we have adopted a sporadic model for star formation,
which utilizes highly localized, randomly distributed bursts of star formation that
inject metals into the interstellar medium. We have considered several 
model galaxies that have masses and rotation curves typical for blue LSB galaxies 
and are characterized by distinct SFRs and IMFs.
We seek to find chemical and photometric signatures 
(such as radial oxygen abundance fluctuations, time evolution of mean oxygen abundances, 
integrated $B-V$ colors and H$\alpha$ equivalent 
widths) that can help to constrain the age of blue LSB galaxies.
We find the following.
\begin{itemize}
\item Local bursts of  star formation leave signatures in the disk in the form of oxygen 
enriched patches, which are subject to a continuing destructive influence of 
differential rotation, tidal forcing from spiral stellar density waves, and 
supernova shock waves from neighboring star formation sites. As a result, our model galaxy 
reveals short scale ($\sim$ 1--2~kpc) radial fluctuations in the oxygen abundance [O/H].
Typical amplitudes of these fluctuations are 0.5--1.0~dex in (1--2)-Gyr-old galaxies
but they decrease quickly to just 0.1--0.2~dex in (5--13)-Gyr-old ones.
\item The oxygen abundance fluctuation spectrum (i.e., the normalized number 
of fluctuations versus the fluctuation amplitude) has a characteristic shallow slope
in (1--2)-Gyr-old galaxies. The fluctuation spectrum is age-dependent and steepens with time, 
reflecting the ongoing mixing of heavy elements in the disk. 
The dependence of the fluctuation spectrum on the model parameters and 
spatial resolution makes it possible to use the spectrum only 
for {\it order-of-magnitude} age estimates.
\item 
The mean oxygen abundance  $\langle {\rm [O/H]} \rangle$ versus 
$EW({\rm H}\alpha)$ diagram can be used to constrain the age of blue LSB galaxies, if some 
independent knowledge of the IMF is available.
\item Numerical models with Salpeter and Kroupa IMFs and standard mass limits (0.1--100~$M_\odot$)
populate the upper-right portion of the observed $\langle {\rm [O/H]} \rangle$ vs. $EW({\rm H}\alpha)$
phase space.
This implies that blue LSB galaxies are likely to have IMFs with a truncated upper mass limit.
\item Our model data strongly suggest the existence of a minimum age for
blue LSB galaxies. From model $B-V$ colors and mean oxygen abundances 
we infer a minimum age of 1.5--3.0~Gyr.
However, model H$\alpha$ equivalent widths imply a larger value, of order 5--6~Gyr. The latter
value may decrease somewhat if LSB galaxies host IMFs with a truncated upper mass limit.
A failure to observationally detect large oxygen abundance
fluctuations, which, according to our modelling, are typical for (1--2)-Gyr-old 
galaxies, will argue in favour of the more evolved nature of blue LSB galaxies.
\item The oldest galaxies (13 Gyr) in all numerical models have mean oxygen abundances, 
colors and equivalent widths that
lie within the observed range, suggesting that LSB galaxies are not considerably younger than 
their high surface brightness counterparts. 
\item Low metallicities and moderate cooling
render the formation of cold gas with temperature $\la 50$~K inefficient.
This fact implies that most of the cold phase may be in the form of atomic rather 
than molecular hydrogen, though further numerical simulations are needed to make 
firm conclusions.
\item Sporadic star formation yields no radial abundance gradients in the disk
of our model galaxy, 
in agreement with the lack of oxygen abundance gradients in blue LSB galaxies.
\end{itemize}

\acknowledgement
We appreciate the anonymous referee's suggestion to use population synthesis modelling.
We thank Stacy McGaugh for allowing us to reanalyze 
his multicolor CCD image database of LSB galaxies. This work is partly
supported by the Federal Agency of Science and Education 
(project code RNP 2.1.1.3483), by the Federal Agency of Education 
(grant 2.1.1/1937), by the RFBR (project code 06-02-16819) 
and by SFB 591. E. I. V. gratefully acknowledges support from an ACEnet Fellowship.
The numerical simulations were performed on the Atlantic Computational Excellence Network
(ACEnet). We also thank Professor Martin Houde for access to computational facilities.

\appendix 
\section{Disk scale height}
We derive the disk vertical scale height $Z$ at each computational cell 
via the equation of local vertical pressure balance 
\begin{equation}
\rho_{\rm g} c_s^2 = 2\int_0^Z \rho_{\rm g} \left( g_{z}^{\rm gas}+g_{z}^{\rm st} +
g_{z}^{\rm halo} 
\right) dz,
\label{eqn1}
\end{equation}
where $\rho_{\rm g}$ is the gas volume density, $g_{z}^{\rm gas}$ and $g_{z}^{\rm st}$ 
are the {\rm vertical} gravitational accelerations due to self-gravity of gas and stellar layers,
respectively, and $g_{z}^{\rm halo}$ is the gravitational pull of 
a dark matter halo. 
Assuming that $\rho_{\rm g}$ is a slowly varying function of vertical distance $z$ between $z=0$ (midplane)
and $z=Z$ (i.e. $\Sigma_{\rm g}=2\, Z \,\rho_{\rm g}$) and using the Gauss theorem, one can show that
\begin{eqnarray}
&&\int_0^Z \rho_{\rm g} \, g_{z}^{\rm gas} \,  dz = {\pi \over 4} G \Sigma_{\rm g}^2 \label{eq2} \\
&& \int_0^Z \rho_{\rm g} \, g_{z}^{\rm st} \,  dz = {\pi \over 4} G \Sigma_{\rm st} \Sigma_{\rm g} \label{eq3} \\
&& \int_0^Z \rho_{\rm g} \, g_{z}^{\rm halo} \, dz = {G M_{\rm h}(r) \rho_{\rm g} \over r} 
\left\{  1-\left[ 1+ \left({\Sigma_{\rm g} \over 2 \rho_{\rm g} r} \right) \right]^{-1/2} \right\},
\label{eq4}
\end{eqnarray}
where $r$ is the radial distance and $M_{h}(r)$ is the mass of the dark matter halo located inside 
radius $r$. In equation~(\ref{eq3}) we have assumed that the volume density of stars can be written
as $\Sigma_{\rm st}/Z$.
Substituting equations~(\ref{eq2})-(\ref{eq4}) back into equation~(\ref{eqn1}) we obtain
\begin{equation}
\rho_{\rm g} \, c_s^2 = {\pi \over 2} G \Sigma_{\rm g} + {\pi \over 2} G \Sigma_{\rm g} \Sigma_{\rm
st} + {2 G M_{\rm h}(r) \rho_{\rm g} \over r} 
\left\{  1-\left[ 1+ \left({\Sigma_{\rm g}  \over 2 \rho_{\rm g} r} \right) \right]^{-1/2} \right\}.
\label{height1}
\end{equation}
This can be solved for $\rho_{\rm g}$ given the model's known $c_s^2$, $\Sigma_{\rm g}$, $\Sigma_{\rm
st}$, and $M_{\rm h}(r)$.
The vertical scale height is finally derived as $Z=\Sigma_{\rm g}/(2\rho_{\rm g})$.

\section{Population synthesis modelling}

We make use of the single-burst models by \citet{bc03} to synthesize 
broad-band colors and 
H$\alpha$ equivalent widths. The continuum level $F_{\rm c}(t)$ near the $H\alpha$ line
at time $t$ can be found as 
\begin{equation}
\label{bc1}
F_{\rm c}(t) = \int_0^{t} f_{\rm c}(t-\tau,Z_m)\, \Psi(\tau)\, d\tau ~,
\end{equation}
where $\Psi(\tau)$ is the SFR at time $\tau$, and $f_c(t-\tau, Z_{\rm m})$ is 
the continuum luminosity per wavelength for a single-burst stellar population
with age $t-\tau$ and metallicity $Z_{\rm m}$.

The H$\alpha$ luminosity $L_{{\rm H}\alpha}(t)$ at time $t$ 
is estimated as $L_{H\alpha}\, [\mathrm{erg/s}] = 1.36 \times 10^{-12} N_{\rm Lyc} 
[\mathrm{1/s}]$  \citep{kennicutt83},
where $N_{\rm Lyc}$ is the rate of ionizing photon production derived directly
from Bruzual and Charlot's population synthesis modelling.
The resulting equivalent width is found 
as $EW({\rm H}\alpha) = L_{{\rm H}\alpha}/F_{\rm c}$.
The broad-band $B-V$ color at time $t$ is found 
via $B$ and $V$ luminosities ($L_{\rm B}(t)$ and $L_{\rm V}(t)$, respectively) 
calculated using equation~(\ref{bc1}), in which $f_{\rm c}$ is replaced with luminosities in 
the $B$ and $V$ bands, respectively. The resulting color is 
$(B-V) \,=\, -2.5\log_{10}[L_{\rm B}(t) \,/\, L_{\rm V}(t)]$.

\begin{figure}
  \resizebox{\hsize}{!}{\includegraphics{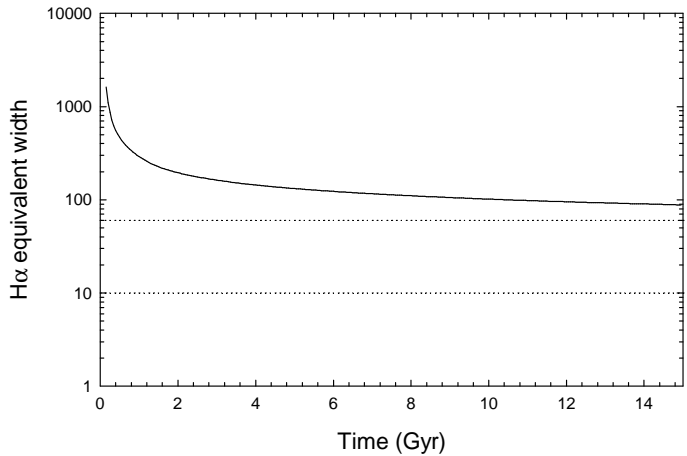}}
      \caption{H$\alpha$ equivalent width versus galactic age 
      for a test model as described in the text. The solid line
      presents the data for metallicity $Z_{\rm m}=0.001$. The horizontal
      dotted lines delineate the observed range of values for blue LSB galaxies.}
         \label{fig12}
\end{figure}

We perform a consistency test on our population synthesis model 
by comparing our predicted $EW({\rm H}\alpha)$ with those
of \citet{zack06} for a model galaxy with a constant (over 15 Gyr) SFR, 
total mass of formed stars $10^{10}~M_\odot$, metallicity $Z_{\rm m}=0.001$,
and Salpeter IMF with mass limits $(m_{\rm low},m_{\rm up})=(0.1,120)~M_\odot$.
The resulting H$\alpha$ equivalent width as a function of time is shown in
Fig.~\ref{fig12} by the solid line.
Our model H$\alpha$ equivalent widths match those of \citet{zack06} quite
well---at $t=15$~Gyr we find $EW({\rm H}\alpha) = 88$~\AA,
whereas figure 2 in \citet{zack06} shows $EW({\rm H}\alpha)\approx 86$~\AA.
This example demonstrates that a constant SFR produces too large H$\alpha$ equivalent
widths and hence is unlikely.


Finally, we would like to note that $Z_{\rm m}=0.001$ (or $[{\rm O/H}]=-1.3$)
is not typical for present-day blue LSB galaxies. According to \citet{deBlok98} 
and \citet{Kuzio04}, 
the mean oxygen abundance of blue LSB galaxies is $\langle{\rm [O/H]}\rangle_{\rm LSB}=-0.85$~dex or
one seventh of the solar metallicity. Higher metallicity is expected to yield 
lower H$\alpha$ equivalent widths.

\end{document}